\documentclass[showpacs,aps,prb,reprint,superscriptaddress,preprintnumbers]{revtex4-1}

\usepackage{amsmath}
\usepackage{amssymb}
\usepackage{graphicx}% Include figure files
\usepackage{dcolumn}
\usepackage{color}
\usepackage{xcolor}
\usepackage{endnotes}
\usepackage{bm}
\usepackage{epsfig}
\usepackage{array}

\newcommand{\etal}{{\it et al.}}

\begin{document}

%\preprint{v3.1}

\title{Angular dependence of the upper critical field in the high-pressure $1T'$ phase of MoTe$_2$}

\author{Y. J. Hu$^\S$}
\author{Yuk Tai Chan$^\S$}
\author{Kwing To Lai}
\email[]{ktlai@phy.cuhk.edu.hk}
\author{Kin On Ho}
\author{Xiaoyu Guo}
\affiliation{Department of Physics, The Chinese University of Hong Kong, Shatin, Hong Kong}
\author{Hai-Peng Sun}
\affiliation{Shenzhen Institute for Quantum Science and Engineering and Department of Physics, Southern University of Science and Technology, Shenzhen 518055, China}
\affiliation{Shenzhen Key Laboratory of Quantum Science and Engineering, Shenzhen 518055, China}
\affiliation{Department of Physics, Harbin Institute of Technology, Harbin 150001, China}
\author{K.~Y.~Yip}
\author{Dickon~H.~L.~Ng}
\affiliation{Department of Physics, The Chinese University of Hong Kong, Shatin, Hong Kong}
\author{Hai-Zhou Lu}
\affiliation{Shenzhen Institute for Quantum Science and Engineering and Department of Physics, Southern University of Science and Technology, Shenzhen 518055, China}
\affiliation{Shenzhen Key Laboratory of Quantum Science and Engineering, Shenzhen 518055, China}
\author{Swee K. Goh}
\email[]{skgoh@cuhk.edu.hk}
\affiliation{Department of Physics, The Chinese University of Hong Kong, Shatin, Hong Kong}

\date{\today}

\begin{abstract}
Superconductivity in the type-II Weyl semimetal candidate MoTe$_2$ has attracted much attention due to the possible realization of topological superconductivity. Under applied pressure, the superconducting transition temperature is significantly enhanced, while the structural transition from the high-temperature 1$T'$ phase to the low-temperature $T_d$ phase is suppressed. Hence, applying pressure allows us to investigate the dimensionality of superconductivity in 1$T'$-MoTe$_2$. We have performed a detailed study of the magnetotransport properties and upper critical field $H_{c2}$ of MoTe$_2$ under pressure. The magnetoresistance (MR) and Hall coefficient of MoTe$_2$ are found to be decreasing with increasing pressure. In addition, the Kohler's scalings for the MR data above $\sim$11~kbar show a change of exponent whereas the data at lower pressure can be well scaled with a single exponent. These results are suggestive of a Fermi surface reconstruction when the structure changes from the $T_d$ to 1$T'$ phase. The $H_{c2}$-temperature phase diagram constructed at 15 kbar, with $H\parallel ab$ and $H\perp ab$, can be satisfactorily described by the Werthamer-Helfand-Hohenberg model with the Maki parameters $\alpha \sim$ 0.77 and 0.45, respectively. The relatively large $\alpha$ may stem from a small Fermi surface and a large effective mass of semimetallic MoTe$_2$. The angular dependence of $H_{c2}$ at 15 kbar can be well fitted by the Tinkham model, suggesting the two-dimensional nature of superconductivity in the high-pressure 1$T'$ phase. 
\end{abstract}

%\pacs{Preliminary draft, not all references included}
%\keywords{}

\maketitle
\section{Introduction}
Transition metal dichalcogenides WTe$_2$ and MoTe$_2$ have recently been intensively studied owing to their intriguing physical properties \cite{Yan2017}. For example, extremely large magnetoresistance (MR) has been reported in both WTe$_2$ \cite{Ali2014} and MoTe$_2$ \cite{Keum2015}. Further interests are generated when they are considered as candidates of type-II Weyl semimetals \cite{Soluyanov2015,Wu2016prb,Deng2016,Jiang2017nc}, which would have a pair of topologically non-trivial Weyl points at the boundary of electron and hole Fermi surfaces. A recent focus on these materials concerns their superconductivity because this opens up the possibility of finding topological superconductivity, which could stabilize exotic Majorana fermions \cite{Sato2017}. These features are promising for the development of spintronics devices.

Both WTe$_2$ and MoTe$_2$ consist of weakly bonded (W/Mo)-Te layers stacked along the $c$-axis. While WTe$_2$ crystallizes in a noncentrosymmetric orthorhombic $T_d$ phase (space goup: $Pmn2_1$) at ambient pressure, MoTe$_2$ undergoes a first-order structural transition from a centrosymmetric monoclinic $1T'$ phase (space group: $P2_1/m$) to the $T_d$ phase at $T_s \sim$ 250~K. At low temperature, a superconducting phase transition can additionally be observed at $T_c \sim$ 0.1 K \cite{Qi2016}. In contrast, superconductivity in the bulk WTe$_2$ can only be stabilized at high pressure $\gtrsim$25~kbar \cite{Pan2015,Kang2015, Chan2017}. 

An interesting interplay between structural and superconducting transitions in MoTe$_2$ is revealed upon the application of hydrostatic pressure: $T_s$ can be suppressed to zero at $\sim$ 10 kbar, i.e. at high pressure, the $T_d$ phase can be completely removed and the $1T'$ phase takes over. Meanwhile, $T_c$ is rapidly enhanced, leading to a 30-fold increase in $T_c$ ($\sim$4 K) at $\sim$15 kbar \cite{Qi2016,Takahashi2017,Heikes2018,Lee2018,Guguchia2017}.
A similar enhancement of $T_c$ can also be observed in S-, Se- and Re-doped MoTe$_2$ as well as Te-deficient MoTe$_2$, but $T_s$ is only slightly suppressed before suddenly vanishes with increasing doping/deficiency levels \cite{Takahashi2017,Chen2016,Cho2017,Mandal2018}. Therefore, pressurized MoTe$_2$ presents an opportunity to study the nature of the superconductivity in the $1T'$ phase.

%Many efforts have been put to explore the superconductivity in topological semimetals \cite{Butch2011prbrc,Hosur2014prb,Kobayashi2015prl} since its discovery. 
Previous high pressure studies reported the intrinsic superconductivity in many topological materials, including Cd$_3$As$_2$ \cite{He2016qm}, TaAs \cite{Zhou2016prl}, TaP \cite{Li2017qm}, ZrTe$_5$ \cite{Zhou2016pnas}, HfTe$_5$ \cite{Qi2016prb}, TaIrTe$_4$ \cite{Cai2019, Xing2018arxiv} and YPtBi \cite{Meinert2016prl, Butch2011prbrc, Bay2012}. Particularly, the topological semimetal YPtBi has been found to be an unconventional spin-$3/2$ superconductor, which is beyond the value of spin in triplet superconductors \cite{Kim2018scnadv}. 
In MoTe$_2$, the enhanced $T_c$ at high pressure has not been envisaged in previous density functional theory prediction \cite{Riflikova2014prb}. This discrepancy may be due to the 2-dimensional (2D) nature of the superconductivity in MoTe$_2$.
Recently, Heikes \etal\ \cite{Heikes2018} suggested that applying pressure to MoTe$_2$ would induce the decoupling of Mo-Te layers, leading to a more 2D structure. 
If this high-pressure superconducting phase is quasi-2D, it would be a possible route to search for topological superconductivity \cite{Sato2017}. 
Thus, it is desirable to gauge both the anisotropy of the normal state and the superconducting state under pressure. The case of WTe$_2$ is particularly instructive: while its crystal structure is of layered nature and hence highly two-dimensional, the electronic structure and the superconducting state (at $\sim$100 kbar) are practically isotropic. These conclusions for WTe$_2$ are drawn from quantum oscillations \cite{Cai2015,Zhu2015,Wu2017}, angle-resolved photoemission spectroscopy (ARPES) \cite{Wu2017,Sante2017}, and angular dependence of the magnetoresistance \cite{Thoutam2015} for the electronic structure, and the angular dependence of the upper critical field ($H_{c2}$) for the superconducting state \citep{Chan2017}. In this article, we report the anisotropy of the superconductivity in the $1T'$ phase via a measurement of $H_{c2}$ against the field angle down to 30~mK at 15~kbar.

%%%%%%%%%%%%%%%%Figure 1
\begin{figure}[!t]\centering
      \resizebox{9cm}{!}{
              \includegraphics{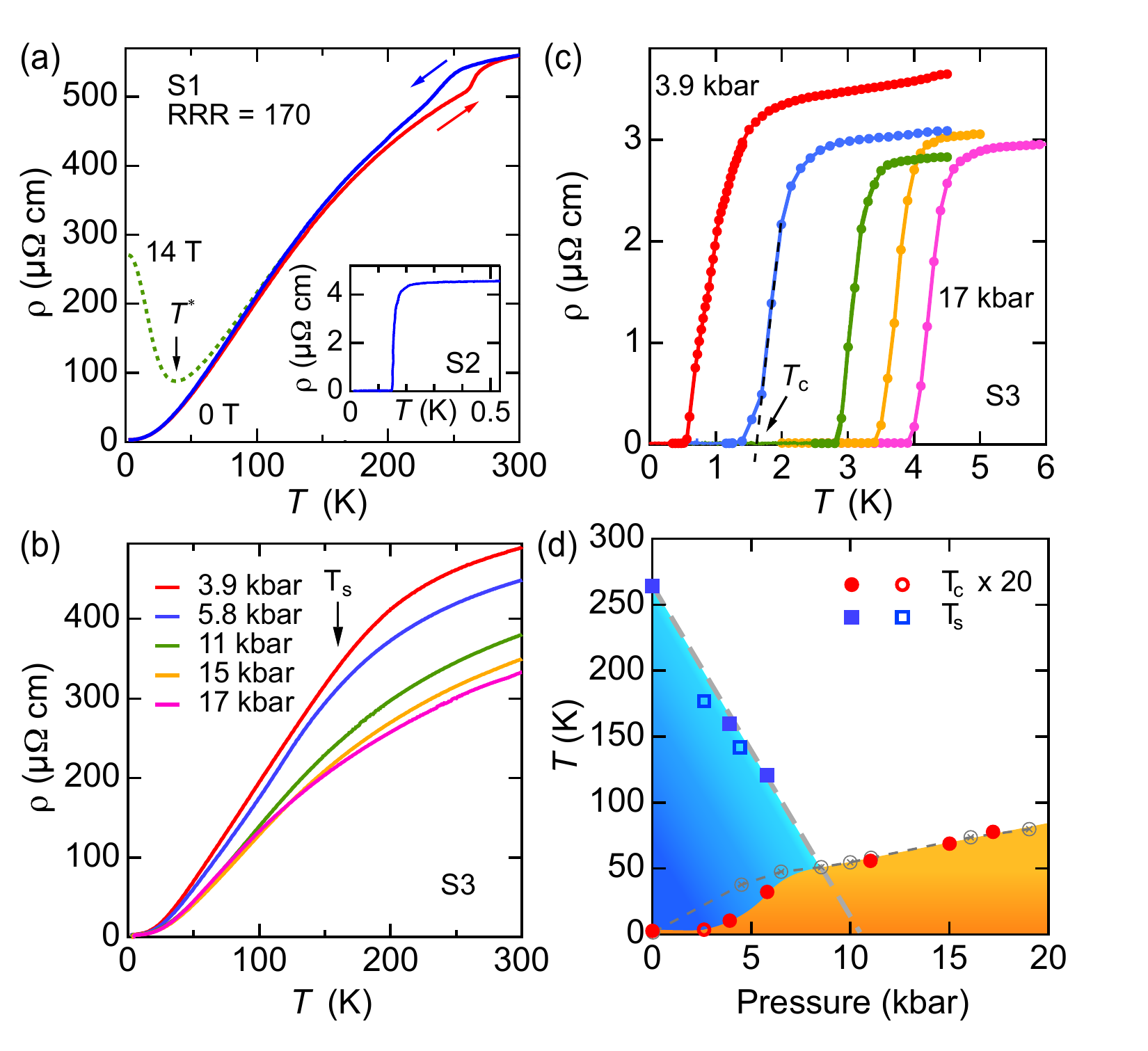}}                				
              \caption{\label{fig1}(a) Temperature dependence of electrical resistivity $\rho(T)$ of MoTe$_2$ (S1) at ambient pressure and zero field (solid curves). The arrows indicate the direction of the temperature sweeps. The dashed curve is $\rho(T)$ at 14~T, showing a field-induced upturn at $T^*$. The inset shows the superconducting transition of S2 at ambient pressure. (b) Pressure dependence of $\rho(T)$ of S3. (c) The superconducting transition of S3 at different pressures. The definition of $T_c$ is indicated. (d) Temperature-pressure phase diagram of MoTe$_2$. The solid symbols represent the data from S3. The open symbols are data from other pressure cell runs. The grey points are data from Ref.~\cite{Guguchia2017}.}
\end{figure}
%%%%%%%%%%%%%%%%%%%%%

\section{Experiment}
Single crystals of MoTe$_2$ were synthesized by the NaCl-flux method as described elsewhere \cite{Keum2015}.  
Temperature dependent electrical transport measurements were performed by a standard four-probe technique in a Bluefors dilution fridge. Hydrostatic pressure dependence was studied by using a piston-cylinder clamp cell with glycerin as the pressure transmitting medium. The pressure value inside the clamp cell was measured by the zero-field superconducting transition of a piece of Pb placed near the sample. 
Magnetic field dependent transport properties were measured with the aid of a superconducting magnet. Transverse resistivity (Hall resistivity) was obtained by symmetrizing (anti-symmetrizing) the field dependent transport data recorded in both positive and negative field directions. 
For the measurements of the angular dependence of $H_{c2}$ at 15~kbar, a miniature moissanite anvil cell was used in conjunction with a vector magnet with a maximum horizontal field of 3~T and a maximum vertical field of 5~T. The pressure  achieved in the anvil cell was determined by ruby fluorescence spectroscopy at room temperature, and glycerin was also used as the pressure transmitting medium. The single crystals used (S1-S4) are from the same growth batch.

\section{Results and discussion}
Figure~\ref{fig1}(a) shows the temperature dependence of the zero-field electrical resistivity $\rho(T)$ (solid lines) of MoTe$_2$ (S1) at ambient pressure. A pronounced anomaly in $\rho(T)$ is recorded at $T_s\approx$ 260~K. This anomaly exhibits a strong hysteresis, signaling a first-order structural transition from $1T'$ to $T_d$ phase, which is consistent with previous reports \cite{Keum2015,Qi2016,Takahashi2017,Lee2018,Heikes2018}. The residual resistivity ratio (RRR) for this sample (S1) is 170, which is a typical value for all samples used in this study. Figure~\ref{fig1}(a) additionally illustrates $\rho(T)$ data at 14~T from 120~K to 2~K. Below $T^*=38$~K, $\rho(T)$ experiences a large enhancement. Consequently, MR at low temperatures is large and reaches 7956\% at 14~T and 2~K, indicating the existence of highly mobile carriers.

Figures~\ref{fig1}(b) and \ref{fig1}(c) display the zero-field $\rho(T)$ curves under pressure. By increasing pressure, the anomaly associated with $T_s$, as indicated by the arrow, weakens drastically and becomes difficult to discern from 11~kbar. The low-temperature part of $\rho(T)$ shows the evolution of the superconducting transition under pressure (Fig.~\ref{fig1}(c)). The values of $T_c$ are defined as the horizontal intercepts of the straight line extrapolated from the transition region (see the dashed line in Fig.~\ref{fig1}(c)). %$T_c$ increases rapidly with an increasing pressure. 
Figure~\ref{fig1}(d) summarizes the pressure dependence of $T_s$ and $T_c$: upon increasing pressure, $T_s$ decreases and extrapolates linearly to 0 K at 11~kbar while $T_c$ is significantly enhanced. The resultant temperature-pressure phase diagram is generally consistent with previous studies \cite{Qi2016,Takahashi2017,Lee2018,Heikes2018,Guguchia2017}. In particular, zero resistance has been observed in the superconducting state at all pressures investigated (Fig.~\ref{fig1}(c)), in contrast to several reports which covered the same pressure range \cite{Lee2018,Heikes2018}. 

%%%%%%%%%%%%%%%%Figure 2
\begin{figure}[!t]\centering
       \resizebox{9cm}{!}{
              \includegraphics{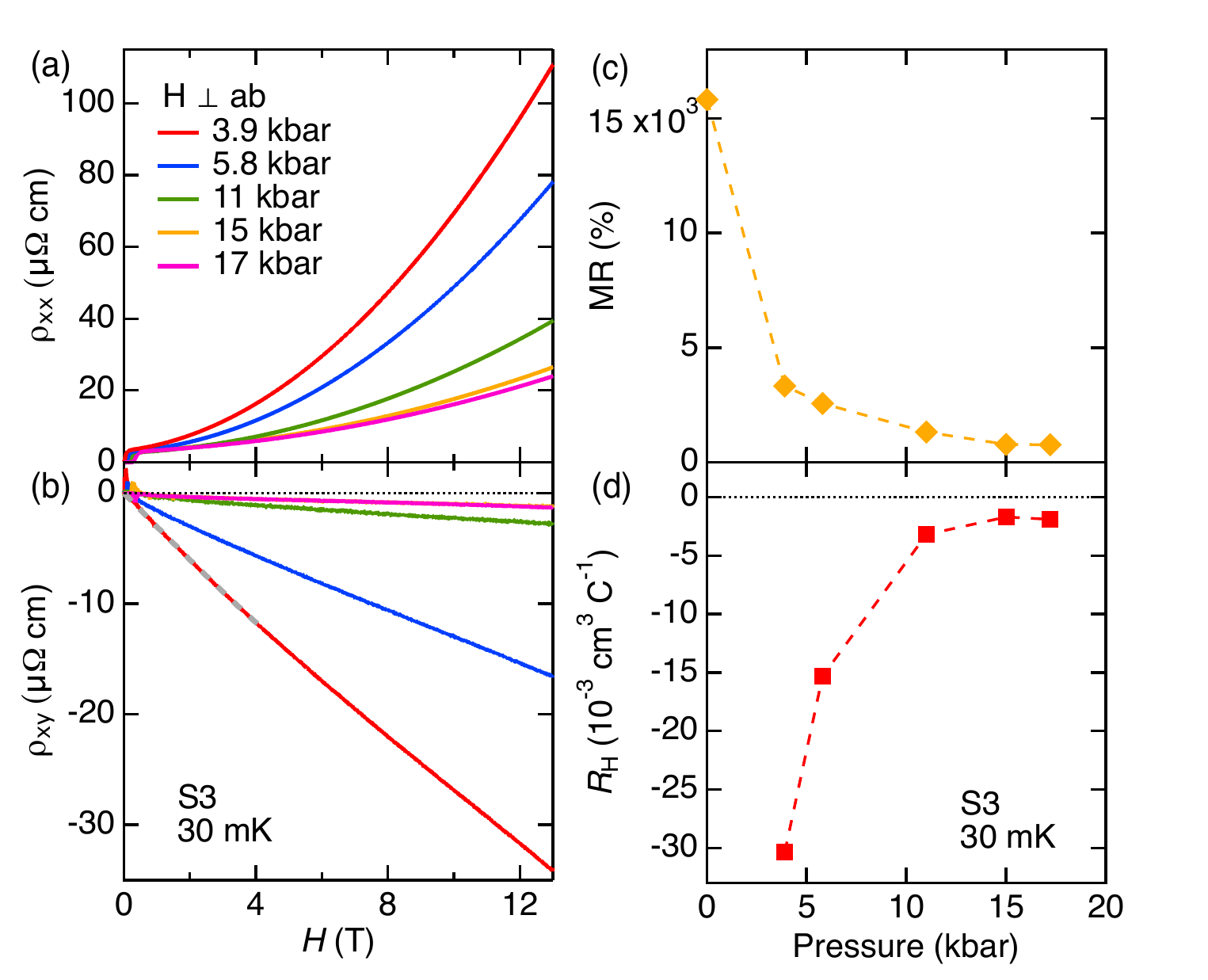}}                				
              \caption{\label{fig2} Field dependence of (a) transverse resistivity $\rho_{xx}$ and (b) Hall resistivity $\rho_{xy}$ for S3 at 30~mK at different pressures. The magnetic field direction is perpendicular to \textit{ab}-plane. The grey dashed line in (b) is the fitting of $\rho_{xy}=R_HH+\beta H^3$ to normal state data below 4~T. (c) Pressure dependence of the magnetoresistance (MR) at 13~T and 30~mK. (d) Pressure dependence of Hall coefficient $R_H$ at 30~mK. 
              %(c,d) Kohler's plots of S3 at 5.8~kbar and 17~kbar, respectively. The dashed lines indicate the $H^2$ dependence.
              }
\end{figure}
%%%%%%%%%%%%%%%%%%%%

In the established temperature-pressure phase diagram, we are able to track the pressure evolution of the electronic structure via magnetotransport. 
Figures~\ref{fig2}(a) and (b) show the field dependence of the transverse resistivity $\rho_{xx}$ and the Hall resistivity $\rho_{xy}$ at 30~mK at different pressures, respectively. The superconducting transition can be seen in both $\rho_{xx}$ and $\rho_{xy}$ at all pressures. At low temperatures, because of the superconducting transition, $\rho_{xx}=0$. Therefore, $\rho_{xx}$(0~T) is extrapolated from the polynomial fitting of the normal state data. $\rho_{xy}$ is determined by first anti-symmetrizing the measured voltage at positive and negative field, and converted by considering the geometry of the sample. The tiny peak at low field, which is close to the superconducting transition, might be experimental artefact and is excluded from the analysis. Figure~\ref{fig2}(c) shows the pressure dependence of MR (=$\Delta\rho_{xx}/\rho(0)$) at 13~T and 30~mK derived from Fig.~\ref{fig2}(a). Figure~\ref{fig2}(d) displays the pressure dependence of the Hall coefficient $R_H$ at 30~mK, which is extracted by fitting the $\rho_{xy}$ data in Fig.~\ref{fig2}(b) with $\rho_{xy}=R_HH+\beta H^3$, where $\beta H^3$ accounts for the small non-linearity in $\rho_{xy}$. Only the normal state data below 4~T are used for this analysis (see the grey dashed line in Fig.~\ref{fig2}(b)).  
When pressure is applied, MR(13~T,~30~mK) first decreases rapidly before levelling off above $\sim$11~kbar, indicating a drastic decrease of carrier mobilities. Meanwhile, a significant initial suppression of $|R_H{\rm (30~mK)}|$ is observed, followed by a nearly constant $|R_H{\rm (30~mK)}|$ above the same pressure ($\sim$11~kbar). $R_H{\rm (30~mK)}$ is negative for all pressure studied, indicating that electrons dominate the electrical transport, while the relative size of electron Fermi pockets increases with pressure. The relatively weak pressure dependence of $|R_H{\rm (30~mK)}|$ and MR(13~T,~30~mK) above $\sim$11~kbar is consistent with the removal of the $T_d$ phase.

%%%%%%%%%%%%%%%%Figure 2.5
\begin{figure}[!t]\centering
       \resizebox{9cm}{!}{
              \includegraphics{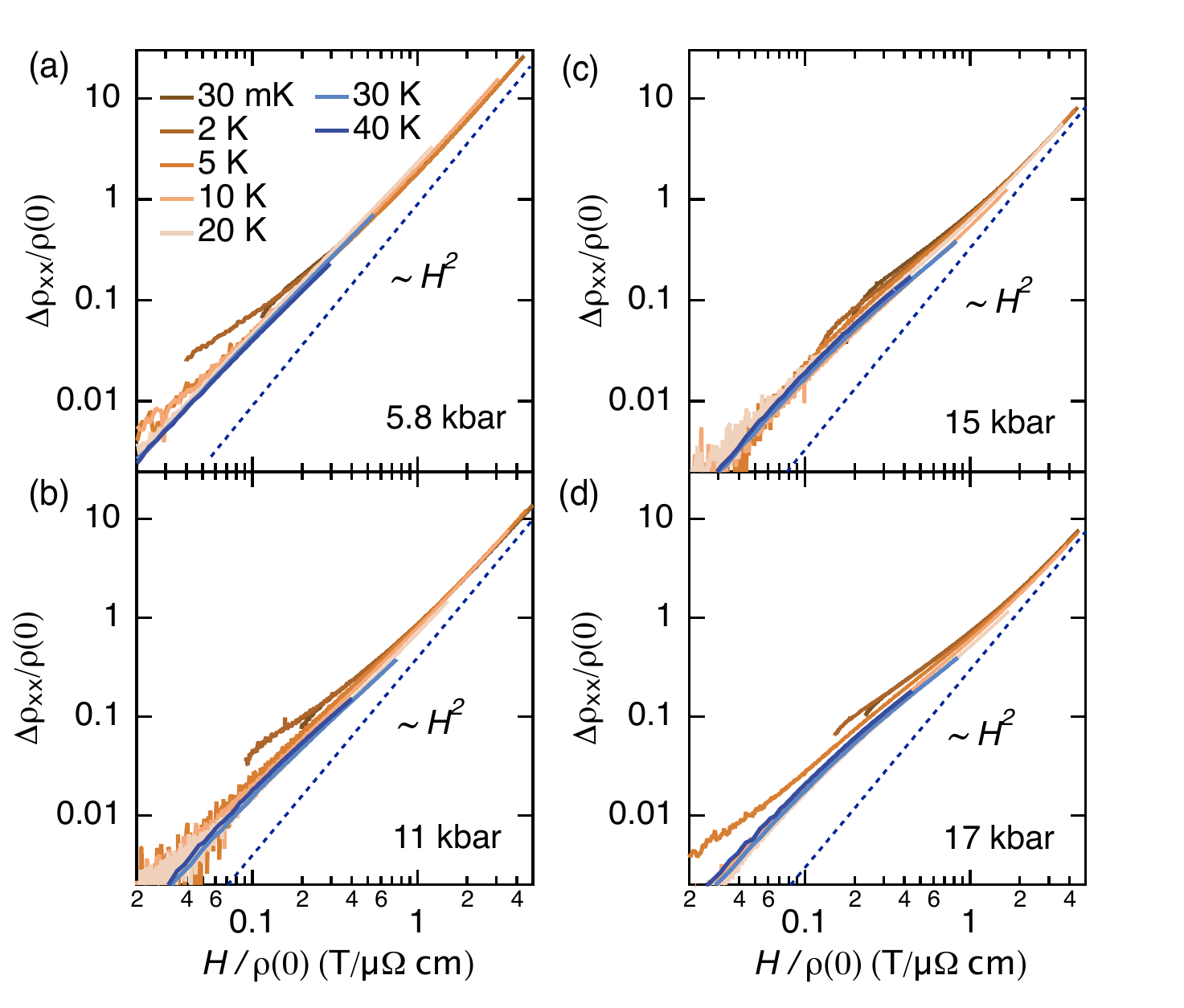}}                				
              \caption{\label{fig2p5} Kohler's plots of S3 at (a) 5.8~kbar, (b) 11~kbar, (c) 15~kbar and (d) 17~kbar. The dashed lines indicate the $H^2$ dependence.}
\end{figure}
%%%%%%%%%%%%%%%%%%%%

Figure~\ref{fig2p5} shows the Kohler plots at 5.8~kbar, 11~kbar, 15~kbar and 17~kbar, respectively. MR against $H/\rho(0)$ is plotted, where $\rho(0)$ is the zero field resistivity at a fixed temperature \cite{supp}. At 5.8~kbar, the data at different temperatures collapse onto a single curve which is nearly quadratic in field, indicating the Kohler's rule is obeyed. The observation of the Kohler's rule has also been demonstrated at ambient pressure \cite{Pei2017}. However, at 15~kbar and 17~kbar, the Kohler's scalings are less satisfied and, when plotted on log-log scales, a slope change is detected. The slope change is also noticeable at 11~kbar (Fig.~\ref{fig2p5}(b)), although the feature is much weaker. This indicates a change in the field exponent and is reminiscent of the case of LaSb \cite{Han2017}, in which a similar change of exponent is noticeable in the Kohler plot at ambient pressure. In LaSb, this behaviour is attributed to the different mobilities associated with different electron Fermi pockets. Thus, if the change of the field exponent detected in MoTe$_2$ at $\geq$11~kbar is similarly rooted on the details of Fermiology, the Fermi surfaces could be different from the ones at $<$11~kbar. This is consistent with the pressure evolution of $|R_H{\rm (30~mK)}|$ and the analysis of Ref.~\cite{Lee2018}, in which they discovered that a four-band model is needed to describe their magnetotransport data above $\sim$~10~kbar, in contrast to the more conventional two-band model applicable for their data at low pressures. The difference of MR between the low and high pressure is again suggestive of the electronic structure reconstruction from the $T_d$ phase to the $1T'$ phase.

%%%%%%%%%%%%%%%%Figure 3
\begin{figure}[!t]\centering
       \resizebox{9cm}{!}{
              \includegraphics{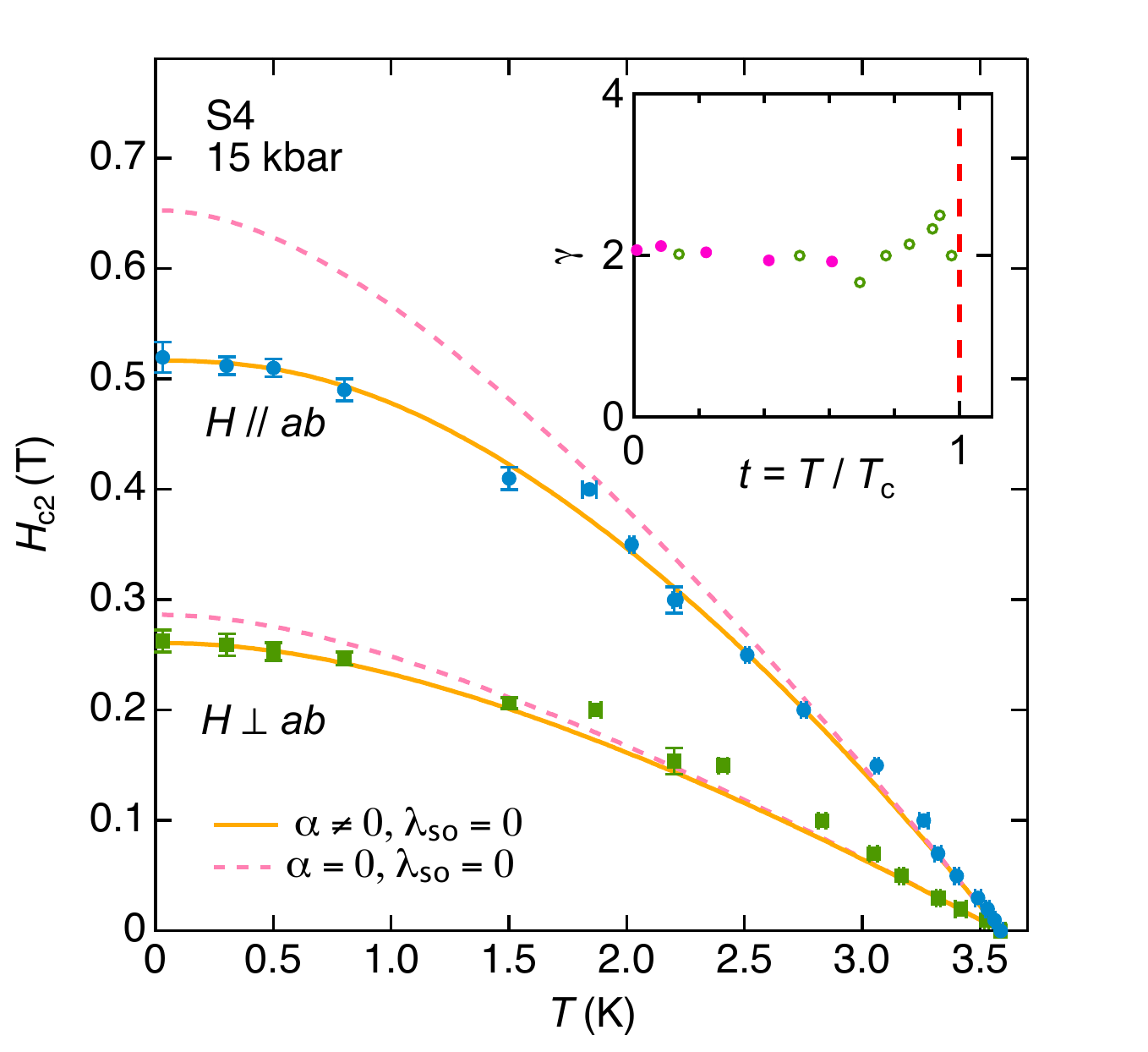}}                				
              \caption{\label{fig3} Field-temperature phase diagram of S4 at 15 kbar, with $H\parallel ab$ and $H\perp ab$. The data (symbols) are measured by both temperature sweeps and field sweeps. The solid and dashed lines are the fits using the WHH model with the Maki parameter $\alpha \neq 0$ and $\alpha = 0$, respectively. The inset displays the temperature dependence of the anisotropy factor $\gamma=H_{c2}(0^\circ)/H_{c2}(90^\circ)$. The solid and open symbols represent the data from the rotation studies and $H_{c2}$ data in the main panel, respectively.}
\end{figure}
%%%%%%%%%%%%%%%%%%%%

Next, we discuss the superconducting state in the high pressure $1T'$ phase.  In the $1T'$ phase, $T_c$ is significantly enhanced, making it easier to investigate the anisotropy of the superconducting state through the measurements of $H_{c2}$ \cite{Chan2017,Chan2018,Goh2012, Shimozawa2014,Naughton1988,He2014,Yonezawa2017,Bay2012}. We have performed the $H_{c2}$ study on MoTe$_2$ (S4) at 15~kbar, which is in the $1T'$ phase according to our phase diagram (see Fig. 1(d)). Figure~\ref{fig3} illustrates the field-temperature phase diagram $H_{c2}(T)$ of MoTe$_2$ at 15~kbar, with $H\parallel ab$ and $H\perp ab$. The raw resistivity data from which these $H_{c2}(T)$ data are determined can be found in the Supplemental Material \cite{supp}.
According to the Werthamer-Helfand-Hohenberg (WHH) theory \cite{Werthamer1966} for a type-II superconductor in the dirty limit, the orbital limited upper critical field is given by 
\begin{equation} \label{orbit}
\left.H_{c2}^{orb}(0)=-0.693\times T_c\dfrac{dH_{c2}}{dT}\right\vert_{T=T_c}.
\end{equation}
The initial slope $(dH_{c2}/dT)_{T=T_c}$ is $-0.26$~T/K and $-0.12$~T/K for $H\parallel ab$ and $H\perp ab$, respectively. Thus, $H_{c2}^{orb}(0)$ are estimated as 0.65~T and 0.29~T, respectively, which are larger than the experimental data at the 0~K limit ($H_{c2}(0)$). The suppression of $H_{c2}(0)$ is more pronounced with $H\parallel ab$. To account for this suppression, we include the Maki parameter $\alpha$. The WHH formula with a finite $\alpha$ is used to fit $H_{c2}(T)$, as displayed in Fig.~\ref{fig3} (solid lines). With $\alpha=0.77$ for $H\parallel ab$ and $\alpha=0.45$ for $H\perp ab$ direction, we are able to describe the $H_{c2}(T)$ data very well.

The Maki parameter $\alpha$ can be written as:
\begin{equation} \label{Maki}
\alpha=\sqrt{2}H_{c2}^{orb}(0)/H_P(0)\sim m^*\Delta(0)/E_F,
\end{equation}
where $H_P(0)$ and $\Delta(0)$ are the Pauli-limiting upper critical field and the magnitude of the superconducting gap at the zero temperature limit, respectively, and $E_F$ is the Fermi energy. Thus, $\alpha$ describes the relative strength of the orbital and spin-paramagnetic (Zeeman) effects. For a conventional metal, $E_F$ is $\sim$1~eV while $\Delta(0)$ is $\sim$1~meV, $\alpha$ is usually much smaller than 1. Therefore, the value of $\alpha=0.77$ is unexpected, indicating a non-negligible spin-paramagnetic contribution to the pair breaking. As stipulated in Equation~\ref{Maki}, an enhanced spin-paramagnetic contribution can come from a small Fermi surface, a large effective mass or a large $\Delta(0)$. Since $T_c$ is low in this system, $\Delta(0)$ alone cannot drive the enhancement of $\alpha$. However, the importance of electron-electron correlation has recently been highlighted \cite{Xu2018, Aryal2019}. Together with the semimetallic nature of MoTe$_2$, the enhancement of $\alpha$ can probably be traced back to the low $E_F$ and high $m^*$. Another possible scenario is that the suppression of $H_{c2}$ could be attributed by the multiband effect with large tunneling between the valleys in Dirac and Weyl semimetals, according to the recent calculation \cite{Rosenstein2018}.

%%%%%%%%%%%%%%%%Figure 4
\begin{figure}[!t]\centering
       \resizebox{9cm}{!}{
              \includegraphics{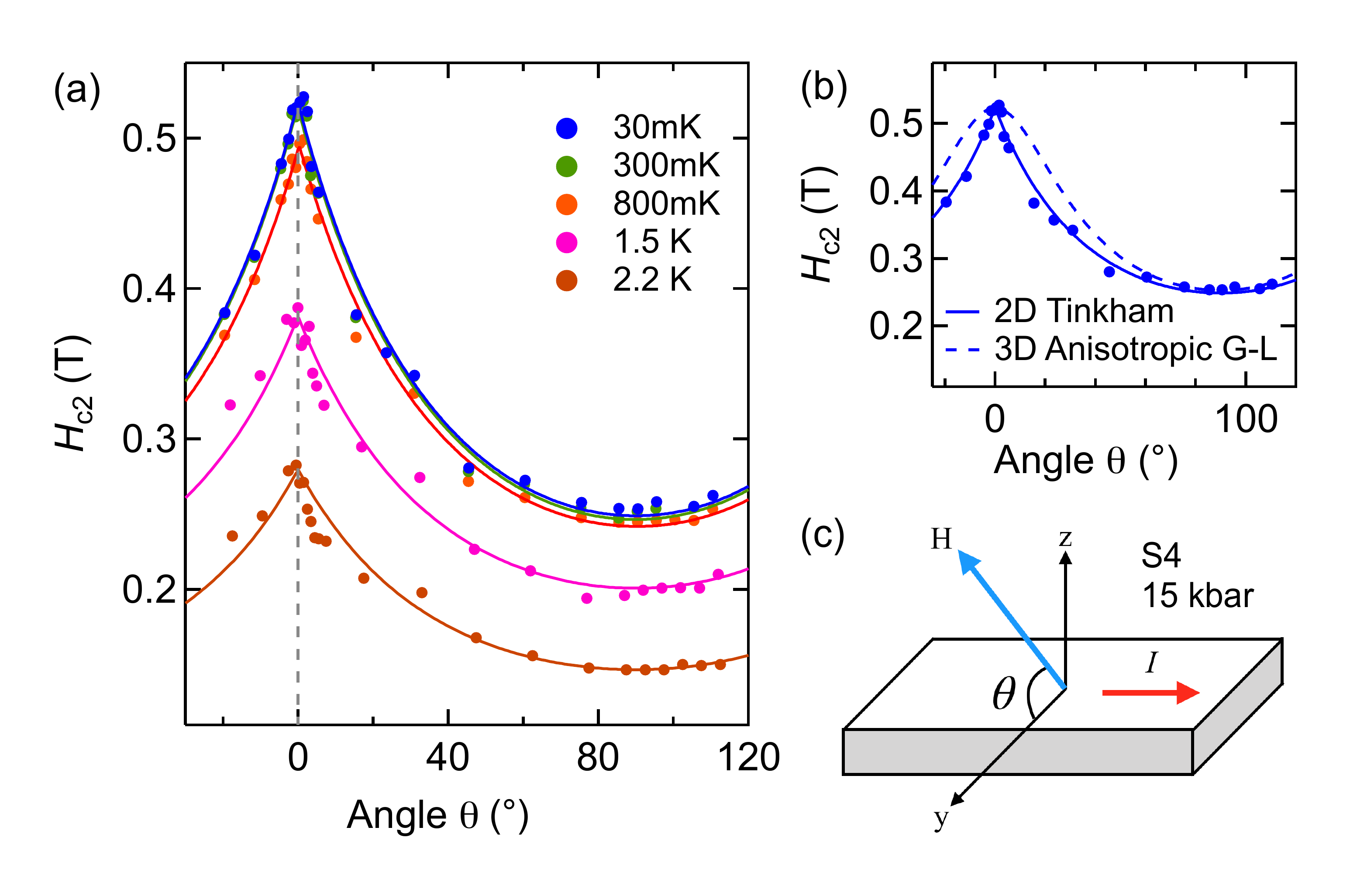}}                				
              \caption{\label{fig4} (a) $H_{c2}(\theta)$ of S4 at 15 kbar at different temperatures. The solid lines are the fits using the Tinkham model. (b) Comparison between the 2D Tinkham model (solid line) and the 3D anisotropic mass model (dashed line) for $H_{c2}(\theta)$ at 30 mK. (c) The arrangement of the sample and the magnetic field. The $z$-axis is parallel to the $c$-axis of the sample, while the $y$-axis lies in the $ab$-plane. The current is always perpendicular to the magnetic field.}
\end{figure}
%%%%%%%%%%%%%%%%%%%%

We now assess the anisotropy of the superconductivity in the $1T'$ phase via a full angular dependence of the upper critical field $H_{c2}(\theta)$ at selected temperatures between 30~mK (0.008$T_c$) and 2.2~K ($0.61T_c$), as illustrated in Fig.~\ref{fig4}(a). The definition of the angle $\theta$ is shown in Fig.~\ref{fig4}(c), where $\theta=0^\circ$ ($90^\circ$) corresponds to $H\parallel ab$ ($H\perp ab$).
At all temperatures studied, $H_{c2}(\theta)$ exhibits a distinct cusp around $H\parallel ab$, which can be well described by the Tinkham model for 2D superconductivity \cite{Tinkham1963}:
\begin{equation} \label{Tinkham}
\left|\dfrac{H_{c2}(\theta)\mathrm{sin}(\theta)}{H_{c2}(90^{\circ})}\right|+\left[\dfrac{H_{c2}(\theta)\mathrm{cos}(\theta)}{H_{c2}(0^{\circ})}\right]^2=1.
\end{equation}
Figure~\ref{fig4}(b) shows the comparison between the 2D Tinkham model and the 3D anisotropic mass Ginzburg-Landau (G-L) model. The 3D anisotropic mass G-L model clearly fails to capture the cusp at $0^{\circ}$. Therefore, the superconductivity in $1T'$ MoTe$_2$ is identified to be 2-dimensional. This is in sharp contrast to the case of WTe$_2$ at 98.5~kbar, in which $H_{c2}(\theta)$ can be described by the 3D anisotropic mass G-L model \cite{Chan2017}.

Despite the success of the Tinkham model in describing $H_{c2}(\theta)$, the anisotropy factor $\gamma=H_{c2}(0^\circ)/H_{c2}(90^\circ)$ is 2.1, which is rather low (inset of Fig.~\ref{fig3}) and only slightly larger than $\gamma$ of 1.7 established in WTe$_2$ \cite{Chan2017}. Furthermore, the in-plane and out-of-plane coherence lengths at the zero temperature limit, $\xi_{\parallel}$ and $\xi_{\perp}$, respectively, can be extracted from the $H_{c2}$ data, giving $\xi_{\parallel}=35.6$~nm and $\xi_{\perp}=17.8$~nm. The value of $\xi_{\perp}$ is much larger than interlayer distance, which is surprising considering the 2D nature of the superconductivity. In fact, the present case is reminiscent to CaAlSi, a superconductor with a MgB$_2$-like structure. In CaAlSi, $H_{c2}(\theta)$ also follows the Tinkham model with a rather low anisotropy factor \cite{Ghosh2003}. There, $\xi_{\perp}$ is also larger than the thickness of the normal layer, and $\gamma$ ranges from $\sim$2 (similar to the present study) at 0.5$T_c$  to $\sim$3.5 at $\sim$0.9$T_c$. The large out-of-plane coherence length for a 2D superconductor remains a puzzle and has to be reconciled in future.

\section{Conclusions}
In summary, we have constructed the temperature-pressure phase diagram of MoTe$_2$ and investigated the anisotropy of superconductivity of the high-pressure $1T'$ phase at 15 kbar. The first-order structural phase transition temperature $T_s$ (from the high-temperature $1T'$ phase to the low-temperature $T_d$ phase) is suppressed with applied pressure and vanishes at $\sim$11 kbar, while the superconducting transition temperature $T_c$ is significantly enhanced. With the application of pressure, the magnetoresistance (MR) and Hall coefficient decrease and saturate to low values at $>$11 kbar. The Kohler scaling can well describe the MR data at all pressures. Meanwhile, a change of exponent is observed at high pressure, suggestive of a Fermi surface reconstruction. Thus, the temperature-pressure phase diagram, together with the magnetotransport measurements, support that the superconductivity at $>$11 kbar is in the $1T'$ phase. Using the Werthamer-Helfand-Hohenberg model with the inclusion of the Maki parameter $\alpha$, the temperature dependence of upper critical field $H_{c2}$ at 15 kbar, obtained at $H\parallel ab$ and $H\perp ab$, can be nicely described with $\alpha$ = 0.77 for $H\parallel ab$ and $\alpha$ = 0.45 for $H\perp ab$. These surprisingly large $\alpha$ indicate the presence of spin-paramagnetic effect. This behaviour may be related to the low Fermi energy in the semimetallic $1T'$-MoTe$_2$, and the large effective mass due to the non-negligible electron-electron correlation. Finally, the angular dependence of $H_{c2}$ can be described by the Tinkham model over a wide temperature range, indicating that the dimensionality of the superconducting state in the high-pressure $1T'$ phase is two-dimensional in nature.

\begin{acknowledgments}
We acknowledge technical support from Qun Niu, and financial support from Research Grants Council of Hong Kong (GRF/14300418, GRF/14301316, GRF/14300117), CUHK Direct Grant (4053223, 4053299), National Natural Science Foundation of China (11504310, 11574127), Guangdong Innovative and Entrepreneurial Research Team Program (2016ZT06D348), National Key R\&D Program (2016YFA0301700) and Science, Technology, and Innovation Commission of Shenzhen Municipality (ZDSYS20170303165926217 and JCYJ20170412152620376).

$^\S$Y.J.H. and Y.T.C. contributed equally to this work.
\end{acknowledgments}

%\bibliography{MoTe2}
%\bibliographystyle{h-physrev2}
%%%%%%%%%%%%%%%%%%%%%%
\providecommand{\noopsort}[1]{}\providecommand{\singleletter}[1]{#1}%

\end{document}